# Design of a non-linear power system stabiliser using the concept of the feedback linearisation based on the back-stepping technique

E. Babaei, S.A.KH. Mozaffari Niapour, M. Tabarraie

Faculty of Electrical and Computer Engineering, University of Tabriz, 51664, Tabriz, Iran
**Email:** e-babaei@tabrizu.ac.ir**,** s.a.kh.mozaffari.niapour@gmail.com, mehrdad.tabarraie@gmail.com

**Abstract:** This study proposes a feedback linearisation based on the back-stepping method with simple implementation and unique design process to design a non-linear controller with a goal of improving both steady-state and transient stability. The proposed method is designed based on a standard third-order model of synchronous generator. A comparison based on simulation is then performed between the proposed method and two conventional control schemes (i.e. conventional power system stabiliser and direct feedback linearisation). The simulation results demonstrate that fast response, robustness, damping, steady-state and transient stability as well as voltage regulation are all achieved satisfactorily.

**Index terms:** Power system stabilizer; excitation controller; direct feedback linearization technique; nonlinear control; backstepping technique; dynamic stability; transient stability.

## 1. Introduction

In the modern power systems, on one hand the systems become more interconnection and hence more complicated and large. On the other hand, considering the steady increase of demand for electricity makes the power systems to operate near their stability margins in order to be more economic. Consequently, the modern power systems are exposed to critics and stability problems more than ever [1-3]. Power system stabilization problems have been dealt with for many years by both control and power systems communities. The main aim of the controller design for a synchronous generator is to supply the loads of the power system with the least interruption under a certain voltage and frequency. Supplementary excitation control system known as PSS is a key tool to improve the low frequency oscillations. The low frequency oscillations are tied with the dynamic and the steady state stability of power systems. In the past two decades a considerable research has been done in the design and application of the PSS. Introduction of lead-lag compensators which are based on the phase compensation in the frequency domain has made the PSS more successful [4]. Control parameters of the PSS are determined using the linearized model of power system. Therefore, to have a satisfactory damping effect over a wide range of operation points, the PSS parameters should be adjusted carefully which is essential to damp both local and inter-area oscillations.

Practically due to frequent variations in the power system's operating conditions it is important to use a precise model of its dynamics and consequently to design a PSS. Therefore, to solve the stability problem of power systems and introduce proper controller when there are uncertainty in parameters it is necessary to recognize resource of these uncertainties. The most common reasons of uncertainty concerning practical power system are:
- load variations, especially load shedding,
- structural changes in the system resulting from changes at generation and transmission facilities,
- changes in the network configuration, especially short circuit and the number of generation units in operation due to reasons such as generation tripping,
- variation of operating conditions of generators.

Considering the above-mentioned causes of uncertainty in parameters, it can be concluded that the practical power systems are highly nonlinear from the view points of configuration and parameters variation with time [5]. Consequently, these systems have frequent variations in different operating conditions. Therefore, in order to achieve good damping effects and guarantee high reliability over a wide operating range, it is worthwhile looking into the possibility of modern nonlinear and linear robust control methods so as to assure a large stability region. Some valuable recent efforts in the field of power system stability using robust linear control have been conducted in [6-10].

Gibbard [7] has focused on a robust PSS design using phase compensation concept and shows that this approach is widely applicable over the practical range of operating conditions and system configurations. Jabr et al. [8] and [9] have addressed methods for solving robust PSS design problems. These methods conduct the locus of eigenvalues that correspond to unstable and poorly damped modes and move them to a satisfactory region in the s-plane. The stabilizer robustness is accounted for in the design problem by simultaneously considering the state-space representations and multivariable root loci corresponding to different operating scenarios. Finally, they proved that these methods guarantee robustness in widely operating conditions. In [10], a power system with finite time delay is considered in signal transmission. Considering this delay in time, a unified Smith predictor (USP) approach is used to formulate the control design problem in the standard mixed sensitivity framework. In this mixed-sensitivity approach of H∞ control design, the controller is applied to ensure robustness, even though contains fast stable poles. These poles often result in numerical instability while solving the problem using linear matrix inequalities (LMIs). In addition, it is shown that by adopting USP formulation, it is possible to overcome these problems. Nevertheless, the disadvantage of such an approach is that it introduces a number of Lyapunov variables which grows quadratically with the system's size. Therefore, the LMI approach involves a large number of extra variables, although the number of the controller's parameters is small.

References [11-13] have presented some techniques to determine the PSS parameters to improve the damping torque. These techniques are based on the concept of the synchronism, damping torque and phase compensation in the frequency domain. This main drawback of these methods is that their application is very difficult if a higher order model of the system is used.

In [14], the application of the linear quadratic regulator (LQR) controller has been presented to design the PSS. Despite its simplicity, this method is sensitive against the variation of system parameters. Moreover, the capabilities of this method is reduces in the presence of disturbances.

In [15], direct feedback linearisation (DFL) method has been presented. In this method, the system is converted to a linear system with variable change and suitable selection of control input signal. Then the system is stabilized by designing state feedback. This method requires the exact identification of the system parameters. Moreover, with the linear controller some advantages of nonlinear controller are not achieved. Furthermore, the controller is sensitive for uncertainties in the system modeling and disturbances.

In [16], optimized $H_\infty$ control theory has been presented to determine the PSS parameters. This method is based on the linearized model of the system around a certain operation point. But the power system is extremely nonlinear.

Reference [17] the multiplied time integral and absolute error is considered as the cost function. The error is the deviation of the rotor speed. This method is based on the repetitive solution techniques which has long computational time.

This paper is devoted to design a nonlinear controller for a single machine connected to an infinite bus, by using the concept of the feedback linearization based on backstepping method. It uses backstepping design process, designs a sequence of virtual systems of relative degree one, reduces relative degree by one by choosing a virtual input, achieves passivity with respect to a virtual output, and the last virtual is used to close feedback loop. The simple and unique design, are the advantages of the proposed method. With the variation of the control system parameters, the desired performance is obtained. Unlike the DFL method, the proposed method does not require a linear controller. Moreover, there is no need to know the exact nonlinear model of the system. Backstepping process of the controller guarantees its robustness against disturbances and uncertainties. The developing of the proposed method to a multi machine system will be considered in the future paper.

Firstly, the dynamic model of power system is described as well as the problem statement. Then the design procedure of the proposed nonlinear method is presented followed by the robustness analysis. The review of conventional PSS with automatic voltage regulator (AVR) and DFL method is then presented. Finally, the simulation results and comparative studies are presented to illustrate the performance of the proposed control scheme and its robustness properties.

## 2. Dynamic model description

Fig. 1 shows a single machine infinite bus (SMIB) power system. This model consists of a generator connected through two parallel transmission lines to a very large network approximated by an infinite bus. The dynamic model of SMIB power system is given by the following third order nonlinear equations [18]:

$$\dot{\delta} = \omega_0 \Delta\omega \qquad (1)$$

$$\Delta\dot{\omega} = -\frac{D}{M}\Delta\omega + \frac{1}{M}(P_m - P_e) \qquad (2)$$

$$\dot{E}_q' = \frac{1}{T_{d0}'}\left[-E_q - (X_d - X_d')I_d + E_{fd}\right] \qquad (3)$$

where, $\delta$ is the power angle of the generator; $\omega_0$ is the base angular frequency of the generator; $\Delta\omega$ is the deviation of the angular speed; $D$ is the damping constant and M is the inertia constant. $P_e$ and $P_m$ are the electrical output power and mechanical input power, respectively. Also, $E_q'$ is $q$ axis transient voltage; $T_{d0}'$ is d axis open circuit time constant; $X_d$ is the equivalent $d$ axis reactance; $X_d'$ is d axis transient reactance; $I_d$ is $d$ axis component of $I$; $E_{fd}$ is d axis field voltage.

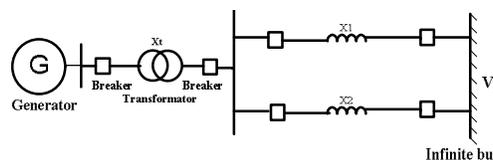

Figure 1. Generator connected through transmission lines to an infinite bus

The equations (1) and (2) show the mechanical dynamics of the generator, and (3) gives the electrical dynamics of the power system. The mechanical input power is treated as a constant in the excitation controller design, such that deviation of the mechanical input power can be neglected or small with respect to the other existing dynamics.
The electrical equations of the generator are expressed as:

$$E_{fd} = K_E u_f \tag{4}$$

$$E_q = E_q' + (X_d - X_d')I_d \tag{5}$$

$$P_e = V_d I_d + V_q I_q \tag{6}$$

For the lossless network, the stator and the network equations are expressed as:

$$V_d = V_B \cos\delta \tag{7}$$

$$V_q = V_B \sin\delta \tag{8}$$

$$E_q' + X_d' I_d = V_{tq} \tag{9}$$

$$X_q I_q = V_{td} \tag{10}$$

$$V_{tq} = V_B \cos\delta + X_e I_d \tag{11}$$

$$V_{td} = V_B \sin\delta - X_e I_q \tag{12}$$

$$V_t = \sqrt{V_{td}^2 + V_{tq}^2} \tag{13}$$

$$E_{fd} = X_{md} I_f \tag{14}$$

In the above equations, $K_E$ is the gain of the excitation amplifier; $E_q$ is $q$ axis voltage; $X_e$ is the equivalent reactance of transmission lines; $V_B$ is the infinite bus voltage; $I_q$ is $q$ axis component of $I$; $X_q$ is $q$ axis reactance; $V_{td}$ and $V_{tq}$ are $d$ and $q$ axis component of $V_t$, respectively; $X_{md}$ is the mutual reactance between excitation coil and stator coil; $I_f$ the is the excitation current.

Considering the above equations, $I_d$ and $I_q$ can be obtained as follows:

$$I_d = \frac{E_q' - V_B \cos\delta}{X_d' + X_e} \tag{15}$$

$$I_q = \frac{V_B \sin\delta}{X_q + X_e} \tag{16}$$

Using (7)-(8) and (15)-(16) in (6), the electrical output power of the generator can be obtained as follows:

$$P_e = \left(\frac{V_B}{X_d' + X_e}\right) E_q' \sin\delta + \frac{V_B^2}{2}\left(\frac{1}{X_q + X_e} - \frac{1}{X_d' + X_e}\right)\sin 2\delta \tag{17}$$

Substituting (17) in (2) also (15) in (3), the system dynamic equations can be derived as follows:

$$\dot{\delta} = \omega_0 \Delta\omega \tag{18}$$

$$\Delta\dot{\omega} = -\frac{D}{M}\Delta\omega + \frac{1}{M}\left(\frac{V_B}{X_d' + X_e}\right) E_q' \sin\delta \\ -\frac{1}{M}\frac{V_B^2}{2}\left(\frac{1}{X_q + X_e} - \frac{1}{X_d' + X_e}\right)\sin 2\delta + \frac{1}{M} P_m \tag{19}$$

$$\dot{E}_q' = \frac{1}{T_{d0}'}(E_{fd} - E_q') + \frac{X_d - X_d'}{T_{d0}'}\left(\frac{E_q' - V_B \cos\delta}{X_d' + X_e}\right) \tag{20}$$

The above equations are used in the controller design in the next section.

## 3. Design of nonlinear excitation controller

Feedback linearization is an effective method to design the nonlinear controller. The advantage of this method is to change the nonlinear dynamic equations of the system to linear equations using the state feedback. Analysis and design of the controllers for the linear systems are simpler and more flexible when compared with the nonlinear systems. However, the transformation of the nonlinear equations to the linear is not unique. As a consequence the results of the design will be different for the different methods of linearization. Moreover, this method has not a satisfactory robustness against disturbances and uncertainties of the model.

The backstepping is a systematic controller design method which composes the selection of the Lyapanov's function with the design of feedback control. This method begins with the differential equations of the systems with the lowest possible order and introduces of the concept of virtual control and designs the desired virtual control step by step. Finally the real control is obtained using the real existing order of the system model.

With the combination of the two mentioned methods (feedback linearization and backstepping technique), the disadvantages of the feedback linearization method can be overcome and a unique design process with easy implementation is obtained.

In this section, the feedback linearization based on the backstepping technique is used which transforms nonlinear system into a linear form by using the backstepping process.

To simplify the dynamic equations the following parameters are defined:

$$\alpha_1 = \omega_0, \quad \alpha_2 = \frac{D}{M}, \quad \alpha_3 = \frac{V_B}{M(X_d' + X_e)}$$

$$\alpha_4 = \frac{V_B}{2M}\left(\frac{1}{X_q + X_e} - \frac{1}{X_d' + X_e}\right)$$

$$\alpha_5 = \frac{1}{T_{d0}'}\left(1 + \frac{X_d - X_d'}{X_d' + X_e}\right), \quad \alpha_6 = \frac{V_B}{T_{d0}'}\left(\frac{X_d - X_d'}{X_d' + X_e}\right)$$

$$\alpha_7 = \frac{P_m}{M}, \quad a = \frac{K_E}{T_{d0}'}$$

The nonlinear power system model, (18)-(20), can be rewritten as follows:

$$\dot{\delta} = \alpha_1 \Delta\omega \tag{21}$$

$$\Delta\dot{\omega} = -\alpha_2 \Delta\omega + \alpha_3 E_{q0}' \sin\delta - \alpha_4 \sin 2\delta + \alpha_7 \tag{22}$$

$$\dot{E}_q' = -\alpha_5 E_q' + \alpha_6 \cos\delta + a u_f \tag{23}$$

In order to apply the proposed method [19], a model of the state deviations around an equilibrium point should be considered as follows:

$$\Delta\delta = \delta - \delta_0 \tag{24}$$

$$\Delta\omega = \omega - \omega_0 \tag{25}$$

$$\Delta E_q' = E_q' - E_{q0}' \tag{26}$$

where, $\Delta\delta$ is the deviation of the power angle; $\delta_0$ is the operating point of power angle; $\Delta E_q'$ is the deviation $q$ axis transient voltage; $E_{q0}'$ is the operating point of $q$ axis transient voltage; $\omega$ is the angular speed of generator.

To stabilize the system in the equilibrium point, the input $u_f$ should have a steady state component. Therefore, it is defined as $u_f = u_{f0} + \Delta u_f$.

Considering (22) and (23) in the steady state and substituting the steady state values of the variables in these equations, $E_{q0}'$ and $u_{f0}$ can be obtained as follows:

$$E_{q0}' = \frac{\alpha_7 - \alpha_4 \sin 2\delta_0}{\alpha_3 \sin\delta_0} \tag{27}$$

$$u_{f0} = \frac{\alpha_5 E_{q0}' - \alpha_6 \cos\delta_0}{a} \tag{28}$$

Assuming that $x_1 = \Delta\delta$, $x_2 = \Delta\omega$, and $x_3 = \Delta E_q'$, the state equations are achieved:

$$\dot{x}_1 = \alpha_1 x_2 \tag{29}$$

$$\dot{x}_2 = -\alpha_2 x_2 - \alpha_3 E_{q0}' \sin(x_1 + \delta_0) - \alpha_4 \sin 2(x_1 + \delta_0) + \alpha_7 - \alpha_3 \sin(x_1 + \delta_0) x_3 \tag{30}$$

$$\dot{x}_3 = -\alpha_5 x_3 - \alpha_5 E_{q0}' + \alpha_6 \cos(x_1 + \delta_0) + a u_{f0} + a \Delta u_f \tag{31}$$

Thus, the system model in the form of usual strictly feedback nonlinear system is as follows:

$$\dot{x}_1 = f_1(x_1) + g_1(x_1) x_2 \tag{32}$$

$$\dot{x}_2 = f_2(x_1, x_2) + g_2(x_1, x_2) x_3 \tag{33}$$

$$\dot{x}_3 = f_3(x_1, x_2, x_3) + g_3(x_1, x_2, x_3) \Delta u_f \tag{34}$$

where

$$f_1(x_1) = 0 \tag{35}$$

$$g_1(x_1) = \alpha_1 \tag{36}$$

$$f_2(x_1, x_2) = -\alpha_2 x_2 - \alpha_3 E_{q0}' \sin(x_1 + \delta_0) \\ - \alpha_4 \sin 2(x_1 + \delta_0) + \alpha_7 \tag{37}$$

$$g_2(x_1, x_2) = -\alpha_3 \sin(x_1 + \delta_0) x_3 \tag{38}$$

$$f_3(x_1, x_2, x_3) = -\alpha_5 x_3 - \alpha_5 E_{q0}' \\ + \alpha_6 \cos(x_1 + \delta_0) + a u_{f0} \tag{39}$$

$$g_3(x_1, x_2, x_3) = a \tag{40}$$

Now, the state equations; (32)-(34), are transformed from $x$ coordinate to the desired $\xi$ coordinate using the proposed method. The design methodology is as follows:

**Step1**:

In this step, $\xi_1$ is defined and then $\xi_2$ is introduced using the backstepping method. Then the differential equation of $\xi_1$ is obtained in terms of $\xi_1$ and $\xi_2$. Introduce $\xi_1 = x_1$, then:

$$a_1(\xi_1) = \lambda_1 \xi_1 + f_1(\xi_1) = \lambda_1 \xi_1 + 0 = \lambda_1 \xi_1 = \lambda_1 x_1 \tag{41}$$

$$\xi_2 = g_1(\xi_1) x_2 + a_1(\xi_1) = \alpha_1 x_2 + \lambda_1 \xi_1 = \alpha_1 x_2 + \lambda_1 x_1 \tag{42}$$

Now, the differential equation of $\xi_1$ can be written as follows:

$$\dot{\xi}_1 = -\lambda_1 \xi_1 + \xi_2 \tag{43}$$

**Step2**:

In this step $\xi_3$ is introduced and the differential equation of $\xi_2$ is obtained. Now introduce

$$\varphi_2(\xi_1, \xi_2) = g_1(\xi_1) f_2(x_1, x_2) \\ = -\alpha_1 \alpha_2 x_2 - \alpha_1 \alpha_4 \sin 2(x_1 + \delta_0) \\ - \alpha_1 \alpha_3 E_{q0}' \sin(x_1 + \delta_0) + \alpha_1 \alpha_7 \tag{44}$$

$$\phi_2(\xi_1, \xi_2) = g_1(\xi_1) g_2(x_1, x_2) \\ = -\alpha_1 \alpha_3 \sin(x_1 + \delta_0) \tag{45}$$

and then

$$\begin{aligned}
a_2(\xi_1,\xi_2) &= \lambda_2\xi_2 + \varphi_2(\xi_1,\xi_2) \\
&+ \left(\frac{da_1(\xi_1)}{d\xi_1} + \frac{dg_1(\xi_1)}{d\xi_1}x_2\right)(-\lambda_1\xi_1 + \xi_2) \\
&= \lambda_1\lambda_2 x_1 + \alpha_1(\lambda_2 + \lambda_1 - \alpha_2)x_2 \\
&- \alpha_1\alpha_3 E_{q0}^{'} \sin(x_1 + \delta_0) \\
&- \alpha_1\alpha_4 \sin 2(x_1 + \delta_0) + \alpha_1\alpha_7
\end{aligned} \quad (46)$$

$$\begin{aligned}
\dot{\xi}_3 &= \varphi_1(\xi_1,\xi_2)x_3 + a_2(\xi_1,\xi_2) \\
&= -\alpha_1\alpha_3 \sin(x_1 + \delta_0)x_3 + \lambda_1\lambda_2 x_1 \\
&+ \alpha(\lambda_2 + \lambda_1 - \alpha_2)x_2 - \alpha_1\alpha_4 \sin 2(x_1 + \delta_0) \\
&- \alpha_1\alpha_3 E_{q0}^{'} \sin(x_1 + \delta_0) + \alpha_1\alpha_7
\end{aligned} \quad (47)$$

Now, the differential equation of $\xi_2$ is calculated as follows:

$$\dot{\xi}_2 = -\lambda_2\xi_3 + \xi_3 \quad (48)$$

**Step3:**
Introduce

$$\begin{aligned}
\varphi_3(\xi_1,\xi_2,\xi_3) &= \phi_2(\xi_1,\xi_2)f_3(x_1,x_2,x_3) \\
&= -\alpha_1\alpha_3 \sin(x_1 + \delta_0)(-\alpha_5 x_3 - \alpha_5 E_{q0}^{'} \\
&+ \alpha_6 \cos(x_1 + \delta_0) + au_0)
\end{aligned} \quad (49)$$

$$\begin{aligned}
\phi_3(\xi_1,\xi_2,\xi_3) &= \phi_2(\xi_1,\xi_2)g_3(x_1,x_2,x_3) \\
&= -a\alpha_1\alpha_3 \sin(x_1 + \delta_0)
\end{aligned} \quad (50)$$

then

$$\begin{aligned}
a_3(\xi_1,\xi_2,\xi_3) &= \lambda_3\xi_3 + \varphi_3(\xi_1,\xi_2,\xi_3) \\
&+ \sum_{j=2}^{3}\left(\frac{\partial\phi_2(\xi_1,\xi_2)}{\partial\xi_{i-1}}x_3\right. \\
&+ \left.\frac{\partial a_2(\xi_1,\xi_2)}{\partial\xi_{j-1}}\right)(-\lambda_{j-1}\xi_{j-1} + \xi_j)
\end{aligned} \quad (51)$$

where

$$\frac{\partial\phi_2(\xi_1,\xi_2)}{\partial\xi_1} = -\alpha_1\alpha_3 \cos(x_1 + \delta_0) \quad (52)$$

$$\frac{\partial\phi_2(\xi_1,\xi_2)}{\partial\xi_2} = -\frac{\alpha_1\alpha_3}{\lambda_1}\cos(x_1 + \delta_0) \quad (53)$$

$$\begin{aligned}
\frac{\partial a_2(\xi_1,\xi_2)}{\partial\xi_1} &= \lambda_1\lambda_2 - 2\alpha_1\alpha_4 \cos 2(x_1 + \delta_0) \\
&- \alpha_1\alpha_3 E_{q0}^{'} \cos(x_1 + \delta_0)
\end{aligned} \quad (54)$$

$$\begin{aligned}
\frac{\partial a_2(\xi_1,\xi_2)}{\partial\xi_2} &= -\frac{2\alpha_1\alpha_4}{\lambda_1}\cos 2(x_1 + \delta_0) \\
&- \frac{\alpha_1\alpha_3}{\lambda_1}E_{q0}^{'}\cos(x_1 + \delta_0) + 2\lambda_2 - \alpha_2 + \lambda_1
\end{aligned} \quad (55)$$

In this step the differential equation of $\xi_3$ is achieved as follows:

$$\dot{\xi}_3 = -\lambda_3\xi_3 + a_3(\xi_1,\xi_2,\xi_3) + \phi_3(\xi_1,\xi_2,\xi_3)\Delta u_f \quad (56)$$

The nonlinear equations (29)-(31) are changed into:

$$\dot{\xi}_1 = -\lambda_1 \xi_1 + \xi_2 \tag{57}$$

$$\dot{\xi}_2 = -\lambda_2 \xi_3 + \xi_3 \tag{58}$$

$$\dot{\xi}_3 = -\lambda_3 \xi_3 + a_3(\xi_1, \xi_2, \xi_3) + \phi_3(\xi_1, \xi_2, \xi_3)\Delta u_f \tag{59}$$

The control law is considered as follows:

$$\begin{aligned}\Delta u_f &= -a_3(\xi_1, \xi_2, \xi_3)/\phi_3(\xi_1, \xi_2, \xi_3) \\ &= \frac{1}{a\alpha_1\alpha_3 \sin(x_1 + \delta_0)} a_3(\xi_1, \xi_2, \xi_3)\end{aligned} \tag{60}$$

The equations of the closed loop system is:

$$\begin{bmatrix}\dot{\xi}_1 \\ \dot{\xi}_2 \\ \dot{\xi}_3\end{bmatrix} = \begin{pmatrix}-\lambda_1 & 1 & 0 \\ 0 & -\lambda_2 & 1 \\ 0 & 0 & -\lambda_3\end{pmatrix}\begin{bmatrix}\xi_1 \\ \xi_2 \\ \xi_3\end{bmatrix} \tag{61}$$

In (61), $\lambda_1$, $\lambda_2$ and $\lambda_3$ are the closed loop poles of the system. It is obvious that when $\lambda_1$, $\lambda_2$, $\lambda_3 > 0$ system (61) is globally stable and it guarantees the exponential convergence of $\xi_{11}$, $\xi_2$, $\xi_3 > 0$ to zero.

As it can be interpreted from (61), the result of the design is a very simple linear closed loop system. In this system, the desired performance is obtained simply by the proper selection of $\lambda_1$, $\lambda_2$ and $\lambda_3$ without any need for linear state feedback controller.

Implementation of the proposed feedback linearization based on backstepping approach requires computation of $\xi_2$ in (42). In order to calculate the value of $\xi_2$, $\varphi_2$ and $\phi_2$ are calculated using (44) and (45), respectively. Afterward by real time computation of partial differentials of $a_1$ and $g_1$ in terms of $\xi_1$ in (46), $a_2$ can be obtained. Then $\xi_3$ is easily achieved using (47). After that $\varphi_3$ and $\phi_3$ are obtained from (49) and (50), respectively. Next making partial differentials of $\varphi_2$ and $a_2$ in terms of $\xi_1$ and $\xi_2$, in (51), $a_3$ is calculated. Finally, the control law is resulted from (60). This nonlinearities term of control law consists of sinusoidal functions and product of variables; however, from the computation complexity aspect; the product of two variables is the same as the product of a constant and a variable. On the other hand, the controller of PSS requires at least measurement of three variables to satisfy inputs of control loop. Considering that in practice the computations should be accomplished on-line, powerful processor is needed for the computations. For this purpose, the fast digital signal processors (DSP) can be used which are commercially available. It is worth noting that the use of DSP for similar on-line complicated mathematical computations has been reported in different references such as [3] and [20], [21]. It is important to note that the experimental implementation of the proposed system is not the aim of this paper and this paper focuses on a new approach for the PSS design.

### 4. Robustness analysis

It should be guaranteed that the feedback control stabilizes the system with respect to its equilibrium point when there is a disturbance or modeling error. The disturbance can be a perturbation in the control input, input mechanical power or measurement of power angle. Modeling errors are as a result of uncertainties in the physical parameters such as uncertainties in the reactance measurements.

The system defined by equations (57)-(59) can be converted to linear close loop system (61) by the law control (60). This result, however, is based on the exact mathematical cancellation of nonlinear terms $a_3$ and $\phi_3$. Exact cancellation is almost impossible for several practical reasons such as model simplification, parameters uncertainty and computational errors. Most likely, the controller will be implementing functions $\hat{a}_3$ and $\hat{\phi}_3$ which are approximations of $a$ and $\phi$, respectively [22]. The actual controller will be implementing the feedback control law as follows:

$$\Delta u_f = -\hat{a}_3(\xi_1, \xi_2, \xi_3)/\hat{\phi}_3(\xi_1, \xi_2, \xi_3) \tag{62}$$

The closed loop system under this control law is:

$$\dot{\xi} = A\xi + B\big(a_3 - \phi_3(-\hat{a}_3/\hat{\phi}_3)\big) \tag{63}$$

The above equation can be rewritten as follows:

$$\dot{\xi} = A\xi + B\delta(\xi) \tag{64}$$

where

$$\delta(\xi) = a_3 - \phi_3 \frac{\hat{a}_3}{\hat{\phi}_3} \tag{65}$$

Thus, the closed loop system appears as the nominal system even when the perturbations are considered.

$$\dot{\xi} = A\xi \tag{66}$$

Since the matrix A is Hurwitz, no serious problem will result from a small error $\delta(\xi)$. To prove the result, it is assumed that $P = P^T > 0$ is the solution of Lyapunov equation

$$PA + A^T P = -Q \tag{67}$$

where $Q = Q^T > 0$.

Suppose that in neighborhood of the region, the error term $\delta(\xi)$ satisfies $\|\delta(\xi)\|_2 \leq \gamma_1 \|\xi\|_2 + \gamma_2$ for some nonnegative constants $\gamma_1$ and $\gamma_2$. This is a reasonable requirement that follows the continuous differentiability form of the nonlinear functions. Using $V(\xi) = \xi^T P \xi$ as a Lyapunov function candidate for the closed loop system, we have

$$\begin{aligned}
\dot{V}(\xi) &= -\xi^T Q \xi + 2\xi^T PB \delta(\xi) \\
&\leq -\lambda_{\min}(Q)\|\xi\|_2^2 + 2\|PB\|_2 \gamma_1 \|\xi\|_2^2 \\
&\quad + 2\|PB\|_2 \gamma_2 \|\xi\|_2
\end{aligned} \tag{68}$$

If $\gamma_1 < \lambda_{\min}(Q)/4\|PB\|_2$, we have:

$$\begin{aligned}
\dot{V}(\xi) &\leq -\frac{1}{2}\lambda_{\min}(Q)\|\xi\|_2^2 + 2\|PB\|_2 \gamma_2 \|\xi\|_2 \\
&\leq -\frac{1}{4}\lambda_{\min}(Q)\|\xi\|_2^2, \quad \forall \ \|\xi\|_2 \geq \frac{8\|PB\|_2 \gamma_2}{\lambda_{\min}(Q)}
\end{aligned} \tag{69}$$

This shows that solutions of the system will be ultimately bounded with an ultimate bound proportional to $\gamma_2$. Furthermore, if $\delta(0) = 0$, can take $\gamma_2 = 0$. In such a case, the origin will be exponentially stable if $\gamma_1 < \lambda_{\min}(Q)/4\|PB\|_2$.

## 5. Review of two conventional methods

In this section, DFL and the conventional PSS with AVR and exciter are explained briefly.

### 5.1. Conventional PSS with AVR

A typical block diagram of an excitation system with AVR and PSS is shown in Fig. 2. It represents a bus-fed thyristor excitation system with an AVR and PSS. The AVR block, as shown in Fig. 2, has been simplified to include only these elements that are considered necessary for operating a specific system. Parameter $T_R$ represents terminal voltage transducer time constant. The exciter is represented by second order dynamic model that consists of an amplifier with derivative feedback. The parameters of exciter are exciter gain, $K_E$, exciter time constant, $T_E$, stabilizer circuit gain, $K_{FE}$, and stabilizer time constant, $T_{FE}$. The nonlinearity associated with the model is that due to ceiling on the exciter output voltage represented by ($E_{fd\max}$, $E_{fd\min}$) and PSS output voltage ($V_{pss\max}$, $V_{pss\min}$). The PSS block consists of a stabilizer gain, $K_{STAB}$, washout time, $T_W$, and two lead-lag compensator with time constant $T_1$ to $T_4$. The stabilizer gain determines the amount of damping introduced by the PSS. The signal washout block serves as a high-pass filter, with the time constant high enough to allow signals association with oscillation in $\omega$ to pass unchanged. Without it, steady changes in speed would modify the terminal voltage. The phase compensation block provides the appropriate phase-lead characteristic to compensate for the phase lag between the exciter input and generator electrical (air-gap) torque. Further information about how to obtain the PSS parameters are available in [23, 24].

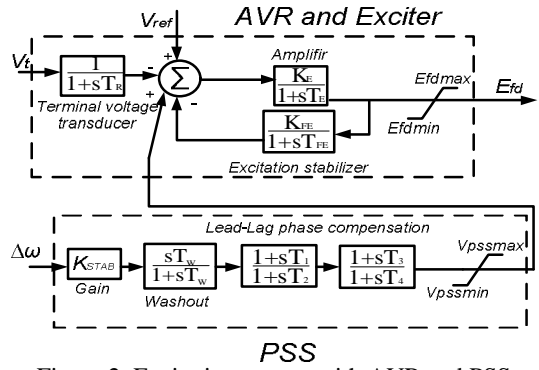

Figure 2. Excitation system with AVR and PSS

*5.2. DFL method*

In this subsection, the design procedure of the feedback linearization is described. As mentioned in the previous sections, in this method a set of state variables are chosen and mapped with a proper mapping function so that the system is linearized. Considering the method in [22], the mapping function, $z = T(x)$, is as follows for the state equations of the system, (29)-(31):

$$z_1 = x_1 \tag{70}$$
$$z_2 = \alpha_1 x_2 \tag{71}$$
$$\begin{aligned} z_3 = &-\alpha_1 \alpha_2 x_2 - \alpha_1 \alpha_4 \sin 2(x_1 + \delta_0) + \alpha_1 \alpha_7 \\ &- \alpha_1 \alpha_3 E_{q0}' \sin(x_1 + \delta_0) - \alpha_1 \alpha_3 x_3 \sin(x_1 + \delta_0) \end{aligned} \tag{72}$$

The state equations of the system are achieved as follows in the new coordinate:

$$\dot{z}_1 = z_2 \tag{73}$$
$$\dot{z}_2 = z_3 \tag{74}$$
$$\dot{z}_3 = \beta^{-1}(x)(\Delta u_f - \alpha(x)) \tag{75}$$

where

$$\beta(x) = -\frac{1}{a\alpha_1\alpha_3 \sin(x_1 + \delta_0)} \tag{76}$$

$$\alpha(x) = -\beta(x)\left( \frac{\partial z_3}{\partial x_1} f_1(x_1) + \frac{\partial z_3}{\partial x_2}(f_2(x_1, x_2) + g_2(x_1, x_2)x_3) + \frac{\partial z_3}{\partial x_3} f_3(x_1, x_2, x_3) \right) \tag{77}$$

Thus the linearization control law is given by:

$$\Delta u_f = \alpha(x) + \beta(x)v \tag{78}$$

It is worth noting that the system equations in the $z$ coordinate are valid in the $0 < x_1 + \delta_0 < \pi$ range. Considering (78), the closed loop system is obtained as follows:

$$\dot{z}_1 = z_2 \tag{79}$$
$$\dot{z}_2 = z_3 \tag{80}$$
$$\dot{z}_3 = v \tag{81}$$

Now to stabilize the system expressed with the above equations, the state feedback $v = -Kz$ is used in which $K = [k_1, k_2, k_3]$ where $k_1, k_2, k_3$ are the state feedback gains. Further information is available in [15].

# 6. Simulation results and comparative studies

The proposed nonlinear control algorithm was simulated in various operating conditions by MATLAB/SIMULINK software. The effectiveness of the proposed method has been validated by comparative studies and simulation results. In this section, the transient and steady state results obtained with proposed approach are compared with the results obtained with two conventional methods. The main circuit is illustrated by the block diagram in Fig. 3 which includes generator, proposed controller, excitation system and AVR system. The system parameters are summarized in Table 1.

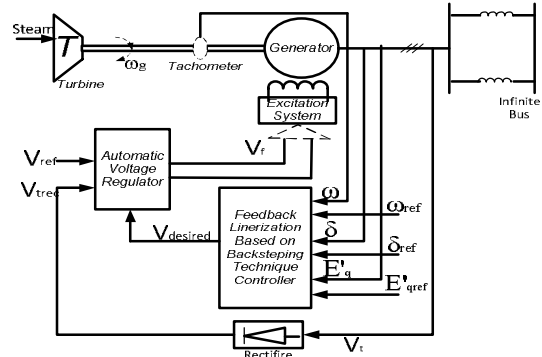

Figure 3. Main circuit schematic of the proposed method

Table 1. Parameters values of the synchronous machine (The base values are 160 MVA rated power and 15 kV rated voltage)

| | | | |
|---|---|---|---|
| $P_o$ | 0.8 pu | $K_{FE}$ | 0.025 |
| $Q_o$ | 0.496 pu | $T_{FE}$ | 1 sec |
| $V_o$ | 1 pu | $E_{fdmax}$ | 4.5 pu |
| $X_d$ | 1.7 pu | $E_{fdmin}$ | -4.5 pu |
| $X_q$ | 1.64 pu | $T_W$ | 6.6 sec |
| $X'_d$ | 0.245 pu | $T_R$ | 0.6 msec |
| $X_e$ | 0.2 pu | $M$ | 6.6 sec |
| $T'_{do}$ | 5.9 sec | $D$ | 0 |
| $K_E$ | 400 | $\omega_o$ | 1 pu |
| $T_E$ | 0.05 sec | $\omega_r$ | 1 pu |

The control parameters have been selected by a fine tuning and finally the following tuned parameters were used in the simulation.

- $K_{STAB}=17.57$, $T_1=1.48s$, $T_2=0.33s$, $T_3=3.55s$, $T_4=11.57s$, $V_{pss\max}=0.15pu$ and $V_{pss\min}=-0.15pu$ for conventional PSS.
- $\lambda_1=5$, $\lambda_2=10$, $\lambda_3=15$ for the proposed method.
- $k_1=1100$, $k_2=185$, $k_3=11$ for DFL controller.

The fault that considered in this paper is a symmetrical three phase short circuit fault exactly on the generator bus bar. The first simulation has been performed to verify the effect of temporary three phase fault by using the following sequence:

- During the time interval 0-0.6 sec, the system is in normal operation.
- At $t=0.6$ sec, a temporary three phase fault occurs.
- At $t=0.78$ sec, the three phase fault is cleared and the system recovers to its initial conditions.

Fig. 4 shows the simulation results for active electric power, power angle, terminal voltage and the relative speed, respectively.

When the fault occurs in the time interval between 0.6 to 0.78 sec, the terminal voltage as well as the active power goes to zero simultaneously. At the same time the power angle and the shaft speed begin to increase. In this condition, the excitation voltage reaches to its maximum value to increase the terminal voltage and electric active power. When the fault is cleared, the rotor speed decreases and the active electrical power reaches to its maximum value. In this situation, in order to avoid the backswing and to reduce the amplitude of the first swing, the control system applies a suitable negative voltage on the excitation system. As a result, the electrical power and the power angle are reduced. This causes the terminal voltage to be reduced. After this process the controller applies a voltage zero to the excitation system. Therefore, the electrical active power, terminal voltage, power angle and the rotor speed recover to their pre-fault steady state values. As Figs. 4(a) and 4(b) show, the power angle and electrical active power reach to their steady state values within only one swing that takes about 1 sec and without any backswing. It is noticeable that the first swing takes place with lower swing amplitude. The other methods have larger swing amplitude and longer settling time. For the DFL and the conventional PSS the settling time is 3 and 5.5 seconds, respectively. It is important to note that the larger settling time might endanger the power system stability. In addition, as shown in Fig. 4(c), in the proposed method the terminal voltage is regulated smoothly with negligible first swing and within about 1 second. While in the DFL method the amplitude of the first swing is not in desired level. Besides, in the conventional PSS method the number of swings and the first swing amplitude is not suitable for the power system. Fig. 4(d) shows that after about 1 sec the rotor speed settles down to its nominal value.

Considering the above discussion and the simulation results, it can be conclude that the proposed controller has a very good performance to improve the transient and steady state stability of the power system under a large disturbance. The results prove the improved capabilities of the proposed controller in the tracking of the power system characteristics.

The second simulation case study has been performed to verify the effectiveness of the proposed control system behavior with respect to 20 percent change of input mechanical power by using the following sequence:

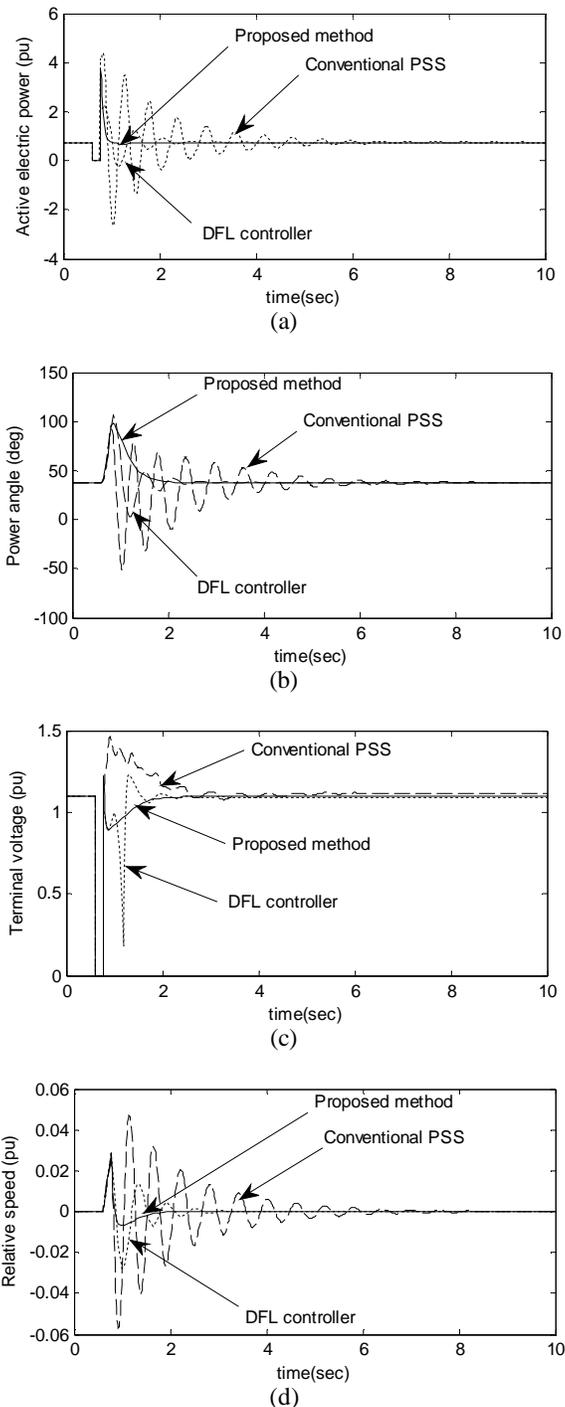

Figure 4. Response of generator in dealing with a temporary three-phase short-circuit fault

(a) Variations of the output active power
(b) Variations of the power angle
(c) Variations of the terminal voltage
(d) Normalized angular speed deviation

- During the time interval 0-1.0 sec, the system is in under normal operation.
- At $t = 1.0$ sec, the mechanical power is increased 20 percent.

Figs. 5(a) and 5(b) show the active electrical output power and the power angle variations as a result of 20% step change in the input mechanical power, respectively. As it can be observed from these figures, the system with the proposed controller has a very good tracking characteristic. Using the proposed controller, the system tracks its new operation point within 0.5 sec and without any overshoot. But with the other methods the first swing of the electrical power reaches to 1.2 pu and the deviation of the power angle from its nominal value is up to 60 degrees. Fig. 5(c) shows the variation of the terminal voltage. Considering this figure, with the proposed method the terminal voltage changes smoothly and quickly to its new value. While with the conventional PSS the terminal voltage has swings and cannot reach to its new operation point. With the DFL method the voltage swing is 0.72 pu that is not in the allowable range of the generator voltage regulation.

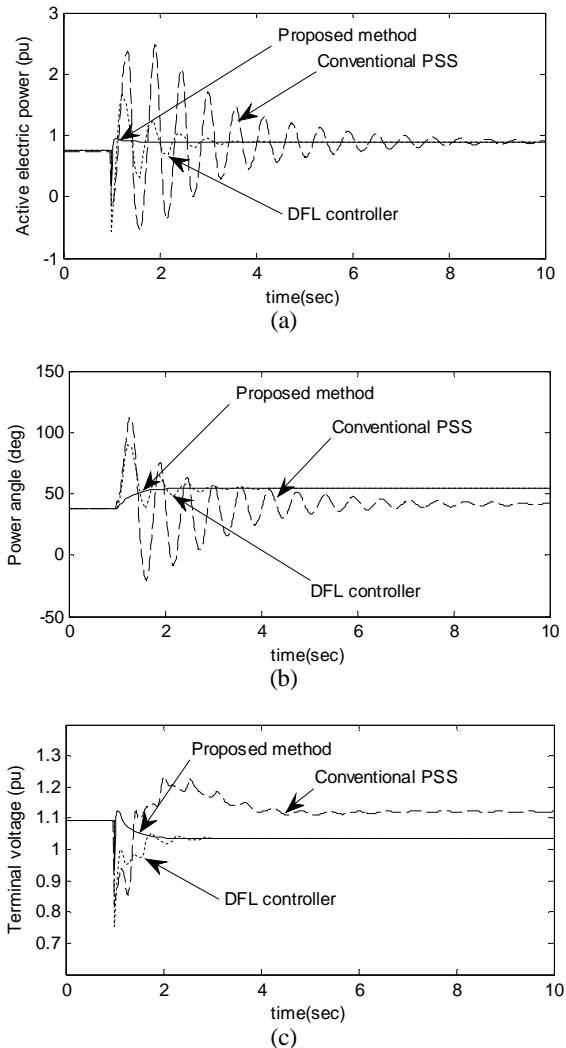

Figure 5. Response of generator in dealing with variation of mechanical input power

(a) Variations of the electrical power arising from a 20 percent variation of mechanical input power
(b) Variations of the power angle arising from a 20 percent variation of mecanical input power
(c) Variations of the terminal voltage arising from a 20 percent variation of mechanical input power

Another simulation case study has been done to investigate the robustness of the system against the duration of three-phase short circuit disturbance and its comparison with the DFL method. The simulation result is shown in Fig. 6(a). As this figure shows the system with the DFL method loses its stability when the duration of the disturbance is longer than 0.869 sec. Considering Fig. 6(b), with the proposed method the system maintains its stability with increase of the short circuit duration up to 0.876 sec. Fig. 6(c) shows the power angle variation of the generator in three different operation points. As this figure shows, the performance of the proposed control and the power system stabilization is independent of the operation points. It is important to note that in the DFL method this is obtained with very high control gain. While in the proposed method the control gain is very low which it is another advantage of the proposed method over the DFL method.

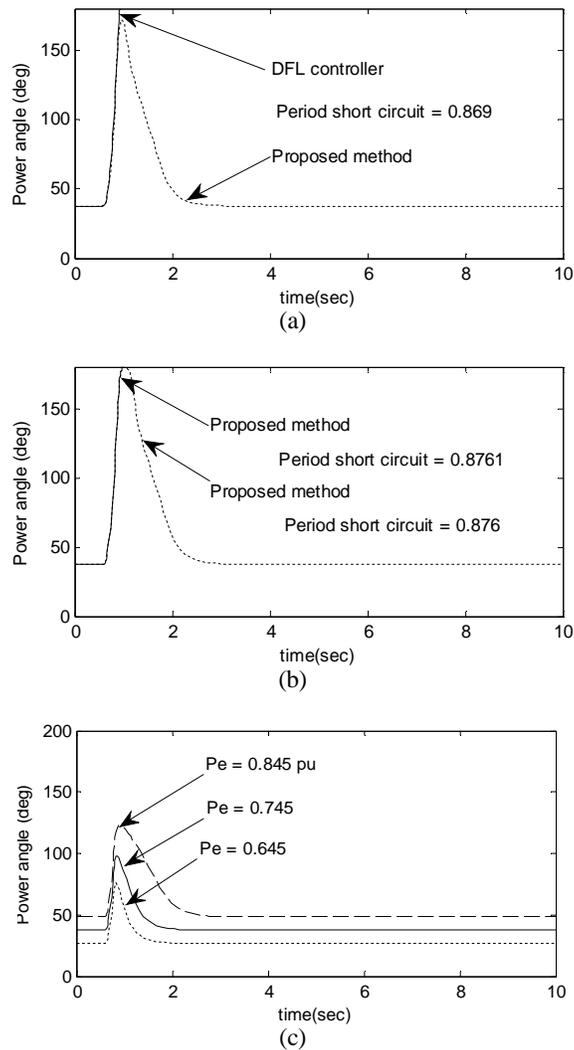

Figure 6. Generator power angle response

(a) Robustness analysis of the proposed method compared to the DFL controller based on the duration of a short circuit
(b) Stability limit analysis of the proposed method based on the duration of a short circuit
(c) Variations of the power angle with the proposed method with respect to the operating point

## 7. Conclusion

This paper proposes a nonlinear power system stabilizer controller with simple implementation and unique design process to achieve both transient and steady state stability enhancement. It represents a realistic alternative to both the conventional PSS scheme and DFL controller. The performance of this controller has been tested through different simulation scenarios and in comparison with the two existing control schemes. As the simulation results verify, the proposed method offers a better performance to recover the system to its nominal conditions after clearance of disturbances and step change in the mechanical input power. The faster response time, more robustness, improved damping and voltage regulation are the main characteristics of the proposed method that result in a more stable operation of the power system from both transient and steady state point of view. Furthermore, the performance of the proposed method is independent of the operating point of the system. The proposed method in this paper could be viewed as a starting point in design and validation of decentralized nonlinear controllers for a multi machine system which will be considered in the future paper from the authors.